\documentclass[aps,showpacs,amsmath,amssymb,prl,superscriptaddress,twocolumn]{revtex4}

\usepackage{epsfig}
\usepackage{amsmath}
\usepackage{amssymb}

\begin{document}

\title{Distribution of Bubble Lengths in DNA}

\author{S. Ares}
\affiliation{Max Planck Institute for the Physics of Complex
Systems,
N\"othnitzer Str. 38, D-01187 Dresden, Germany
and Grupo Interdisciplinar de Sistemas Complejos (GISC)}

\author{G. Kalosakas}
\affiliation{Department of Materials Science, University of Patras,
26504 Patras, Greece}


\begin{abstract}
The distribution of bubble lengths in double-stranded
DNA is presented for segments of varying guanine-cytosine (GC) content,
obtained with Monte Carlo simulations using the Peyrard-Bishop-Dauxois model
at 310 K. An analytical description of the obtained
distribution in the whole regime investigated, i.e., up to
bubble widths of the order of tens of nanometers, is available.
We find that the decay lengths and characteristic exponents of this
distribution show two distinct regimes as a function of GC content.
The observed distribution 
is attributed to the anharmonic interactions within base-pairs.
The results are discussed in the framework of the
Poland-Scheraga and the Peyrard-Bishop
(with linear instead of nonlinear stacking interaction) models.
\end{abstract}
\maketitle

Fluctuating thermal openings of the double helix of DNA (bubbles) are
known to exist in a regime extending well below the denaturation
transition, which includes physiological temperatures. These
openings could be exploited by regulatory proteins or functional
enzymes to perform their work \cite{sobell}. It has been recently suggested that
there is an increased probability for large bubbles to appear at
functionally relevant sites of gene promoter DNA segments
\cite{NAR,EPL}. More accurate numerical calculations \cite{vanerp,rapti}
have unadvertedly reinforced this idea \cite{comm}, by finding an increased
opening probability at functional binding sites of the promoter region
upstream of a gene initiation site.
Such a picture is in analogy to the functional role of fluctuations
in proteins \cite{fenimore}. This has led to an increased interest for
equilibrium and dynamical studies of thermally induced bubbles in DNA.


In a different context, recent
experiments on cyclization of short DNA fragments \cite{cloutier,du}
have drawn attention toward the possibility that regions of very low rigidity
may appear in double stranded DNA \cite{wiggins}. It has been proposed
\cite{yan,ranjith} that these
{\em bending kinks} may be caused by the formation of denaturation bubbles,
letting the higher flexibility of single stranded DNA explain the sharp bending
observed in experiments \cite{cloutier}. In this context, the question of how
frequently denaturation bubbles arise is one of the keys for the
understanding of this phenomenon.

Here we present the distribution of bubble lengths, for
lengths ranging from 0.34 nm (single base-pair openings) up to a
few tens of nanometers (corresponding to openings of size several
tens of base-pairs).
The bubble length distributions have been obtained with
Monte Carlo simulations, using the Peyrard-Bishop-Dauxois (PBD)
model \cite{DPB} for describing the energy of base-pair openings.
 We discuss equilibrium distributions at $T=310$ K,
averaged over DNA segments with total size of one thousand
base-pairs. The bubble length distributions depend on the guanine-cytosine
(GC) content, i.e., the fraction of GC base-pairs in the sequence. The
variation of the distribution as a function of the GC content is
investigated. Note that the size of a bubble is characterized by
its length (width) and its amplitude. Here we fix a threshold
for the amplitude and
when a bubble of length $l$ base-pairs (or $L=0.34 \times l \;$nm)
is mentioned, it means that all the $l$ successive base-pairs have
openings larger than the fixed amplitude threshold $y_{thres}$,
while the limiting
base-pairs (the first neighbors to the left and to the right of the
stretch of the $l$ successive base-pairs) have smaller openings than
that.
Periodic boundary conditions are used, so our study considers only bubbles
in the middle of a sequence, disregarding end effects.
Fraying, i.e. the opening
of the DNA molecule starting at one of its ends, is a very interesting problem
that will be addressed elsewhere.

In this Letter we show that the bubble length distributions,
obtained within the PBD model, follow a
nonexponential law and, further, an analytical description (see Eq.
(\ref{se}) below) is available describing the observed, slower than
exponential, decay with the length. This behavior has been quantified for
various GC contents and remains qualitatively the same independently
on the value of the fixed amplitude. The fact that the same distribution
law is also predicted by the
completely unrelated Poland-Scheraga \cite{poland} model of DNA,
strongly suggests that our findings are beyond theoretical speculation
and might be a proper description of actual DNA physics.

In recent investigations, numerically exact equilibrium properties
\cite{vanerp,rapti} and Langevin dynamics up to nanosecond
time-scales averaged over several hundred realizations
\cite{NAR,EPL,krastan} have been achieved taking advantage of the
efficiency of the simple PBD model for describing
base-pair openings in double-stranded DNA. The PBD model
coarse-grains the relatively rigid internal structure of the
nucleotides and considers their anharmonic stretching interactions
at the single base-pair level \cite{DPB,DP,nlinpeyr}. Its accuracy
has been demonstrated by several comparisons with different
experiments  \cite{CG,NAR,saul}.

The potential energy of the PBD model is given by:
\begin{align}
\label{eq:Hamil}
V=&\sum_n \Big[D_n(e^{-a_ny_n}-1)^2 +\nonumber\\
&\frac{K}{2}(1+\rho e^{-b (y_n+y_{n-1})})(y_n-y_{n-1})^2 \Big].
\end{align}
The sum is over all the base-pairs of the molecule and $y_n$ denotes the relative
displacement from equilibrium at the $n^{th}$ base pair. The first term is an
on-site Morse potential, representing the hydrogen bonds between bases in the
same pair as well as other effective interactions between complementary
nucleotides.
The second term is an anharmonic coupling between adjacent base pairs
that models the stacking interaction. The heterogeneity of the genetic sequence
is taken into account giving different values to the parameters of the Morse
potential for adenine-thymine (AT) or guanine-cytosine (GC) base pairs. The
values of the parameters we have used \cite{params} were fitted in Ref.
\cite{CG} to reproduce thermodynamic properties of DNA. The same parameters
have been subsequently used, without additional fitting procedures,
to successfully describe experimental observations \cite{NAR,saul}.

To study the statistics of bubbles we have performed Monte Carlo simulations of
the PBD model using the same procedure introduced in Ref. \cite{saul}.
The Metropolis algorithm was used to produce equilibrium configurations of the
molecule at $T=310$ K.
Results were averaged, after proper thermalization, over several
realizations (typically 25, each one consisting of $8\cdot10^6$ Monte Carlo
steps, which makes a total of $2\cdot10^8$ steps) with different initial
conditions. As we used quite large DNA sequences (containing 1000
base-pairs) and the temperature studied is well below the
melting temperature, the probability that a complete melting occurred during
the time of a simulation was so low that no melting events (that, for long
enough simulations, would eventually take place for this model at any
temperature \cite{DP,zhang}) were observed.
We show results for bubbles of amplitude equal or greater than
$y_{thres}=1.5$ \AA. We have also studied bubbles of amplitude
$y_{thres}$ 0.5~\AA, 3~\AA and 5~\AA,
finding the same qualitative results presented here. We have checked that
our results are independent of the length of the sequences studied, provided
they are longer than the longest observed bubble and there is no complete
melting.

\begin{figure}
\includegraphics[width=7.9cm]{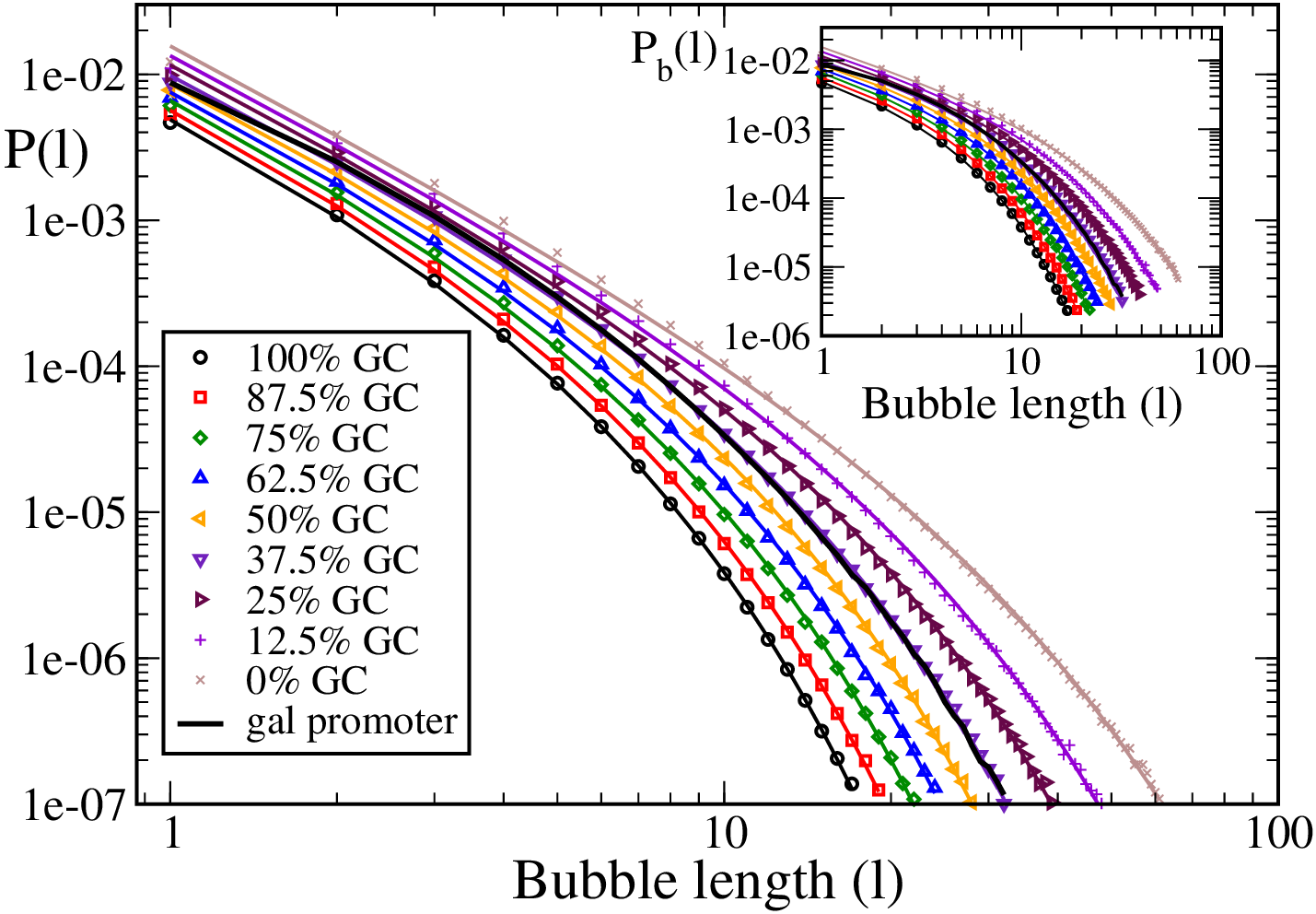}
\caption{\label{fig:1} Distribution per base-pair of bubble sizes $l$ (in
number of base-pairs, $l\ge 1$), $P(l)$, for random sequences with
different GC contents (points, as indicated in the label).
 Also shown is the result for a
segment from {\em E. coli}'s {\it gal} promoter with a 37.45$\%$ GC
percentage (thick line). Thin lines are fits with the analytical
distribution, Eq.(\ref{se}). $T=310$ K, $y_{thres}=1.5$ \AA.
Inset: Probability $P_b(l)=lP(l)$ (for $l \ge 1$) that an individual
baser-pair forms part of a bubble of size $l$ (see text).}
\end{figure}

Figure \ref{fig:1} shows the bubble length distribution
per base-pair, $P(l)$,
obtained from our Monte Carlo simulations. This histogram is
defined as
$P(l)=\lim_{L\to\infty}\langle{N(l)}\rangle/{L}$,
where N(l) is the averaged (during the Monte Carlo simulation)
number of bubbles of length $l$ base pairs on a sequence of total size $L$,
and the average is over different realizations of an uncorrelated
random sequence with a given GC content.
The limit $L\to\infty$ in the definition of $P(l)$ has the
meaning that the total size of the DNA segment should be sufficiently larger
than the considered bubble lengths. Under this condition, the
distribution $P(l)$ should be interpreted as follows: for a given
DNA sequence of total length $L$ base-pairs, the quantity $P(l) \cdot L$
gives the average number of occurrences of a bubble of length $l$ base-pairs,
in thermal equilibrium.

From $P(l)$ we obtain the probability that an individual base-pair
forms part of a
bubble of length $l$ as $P_b(l)~=~lP(l)$ for $l\geq 1$, while the probability
of not belonging to any bubble (i.e., that the base-pair has an opening smaller
than $y_{thres}$) is $P_b(0)=P(0)$. From the definition of $P(l)$ follows
that $P(0)=1-\sum_{l \ge 1} l P(l)$, thus ensuring
correct normalization $\sum_lP_b(l)=1$ of the probability $P_b$.

As a check of our simulations, we have also computed
for pure GC and AT sequences the probability of i) a single
base-pair to have an opening larger than the considered fixed amplitude and
ii) two neighboring base-pairs to simultaneously have openings larger than the fixed
amplitude, using the transfer integral operator
method \cite{ti1,ti2,bsg}. The probabilities calculated
in these two cases, using a different but numerically exact method,
are in a very good agreement (up to numerical accuracy)
with those obtained through the Monte Carlo simulation from the
distributions $P(l)$, as $ \sum_{l \ge 1} lP(l) = 1-P(0)$ and
$\sum_{l \ge 2} (l-1)P(l)$, respectively.

We find that the distributions per base-pair can be
fitted in the whole range of bubble sizes studied \cite{stexp}, with a
single function of the form:
\begin{equation}   \label{se} 
P(l) = W \frac{e^{-l/\xi}}{l^c}, \hspace{0.5cm} \mbox{for} \; l \ge 1. 
\end{equation}
A fit with a stretched exponential function having a stretching exponent
smaller than 1 is also accurate \cite{preprint},
but we have preferred Eq. (\ref{se}) because
this functional form can be derived from existing theories both for a
simplified version of the used model \cite{sung} (the Peyrard-Bishop model,
with linear stacking interactions \cite{comment}) and for the
independent Poland-Scheraga model \cite{poland}.


In order to investigate which interactions of those
present in the model are responsible for
the observed behavior of the
distribution, we have performed similar calculations varying different
terms of the potential of Eq.(\ref{eq:Hamil}). When $\rho=0$, i.e.
the stacking interaction is linearized, the distribution remains
nonexponential, although it is described by smaller $c$, in
complete agreement with previous studies \cite{sung}.
However, when the on-site Morse
potential is linearized (substituting the
Morse potential by its harmonic approximation), then the obtained
distribution is exponential. This happens independently of setting
$\rho=0$ or $\rho=2$. Therefore, we infer that the nonlinear
interactions between the bases within base-pairs result in the
observed bubble length distribution. These anharmonic
on-site interactions have been found to qualitatively affect
other properties of the model as well \cite{charge}. This kind of interaction
appears also in modified models where the stacking interaction
has a qualitatively different shape \cite{joyeux}.

In Figure \ref{fig:1} is also shown, for sake of comparison, a result
for a natural sequence; a segment from {\em Escherichia coli}'s {\it gal}
promoter \cite{genbank}, extending from
the position $-160$ (upstream) up to the site +91 (downstream) around the
transcription start site of {\it galE} gene. This sequence has
a GC fraction of 37.45\%, and, as can be seen from Figure \ref{fig:1}, its
results are indistinguishable from those of a random sequence with almost
the same GC content (37.5\%). This is expected, as it is in agreement
with the known fact \cite{univ} that, despite statistical correlations
in the sequence and local
properties, equilibrium averaged physical properties of large DNA
molecules are basically similar to those of random sequences.

\begin{figure}
\includegraphics[width=8.0cm]{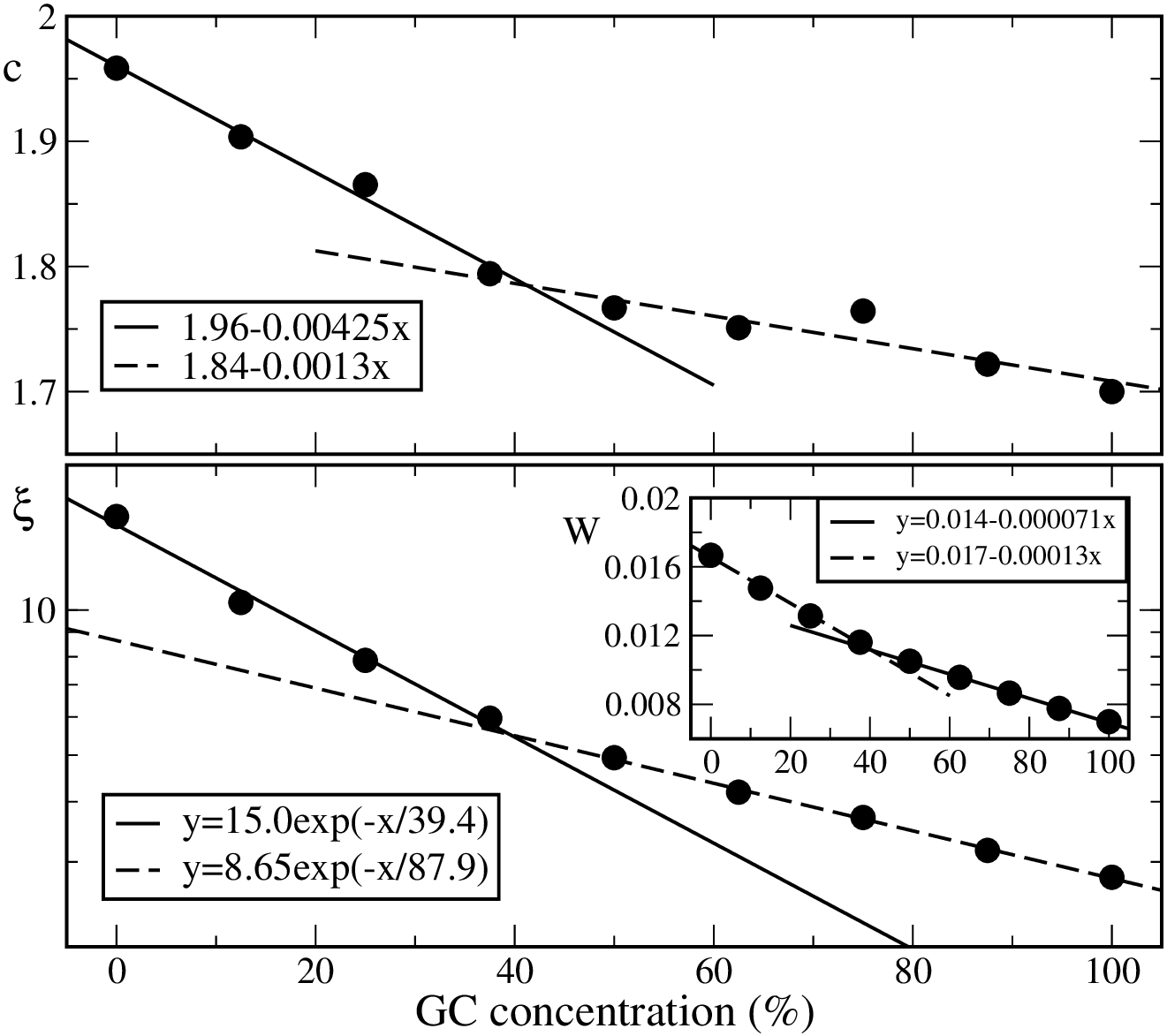} \\ 
\caption{\label{fig:2} Dependence of the exponent $c$
(top), decay length $\xi$ (bottom), and preexponential coefficient
$W$ (bottom, inset) of the distribution, Eq. (\ref{se}), on the GC content
of the sequence (circles). Lines show fits of two distinct regimes with
linear (exponential for $\xi$) functions.
$T=310$ K, $y_{thres}=1.5$ \AA. }
\end{figure}

DNA molecules with higher GC content exhibit a faster decay (see Figure
\ref{fig:1}), as the stronger bound GC pairs make large bubbles less likely.
The change of the distribution with the GC content is
quantified in Figure \ref{fig:2}, where the dependence of the parameters
of Eq. (\ref{se}) on the GC fraction is shown.
All the parameters decrease monotonically with the GC
percentage.
The decay of $W$ signifies that the
higher the GC content the more difficult it is to excite large openings in the
double strand, therefore in AT rich sequences bubbles have a higher statistical
weight.
The decay length $\xi$ is smaller for GC-rich sequences, in which the
distribution decays faster.
The variation of all these parameters can be fitted by a bilinear or
a biexponential function (see Figure \ref{fig:2}), revealing two distinct
regimes on the GC-fraction dependency: above and below a GC content of
about 40\%.
Previous studies \cite{rapti} have shown that the nucleation of bubbles
depends strongly on the sequence: the weaker AT base-pairs have to go over
the potential barrier imposed by their GC neighbors in order to break
the bonds. Our findings suggest that for GC concentrations over 40\%,
the formation of bubbles is a GC-dominated process, as AT base pairs are not
in average free to melt without GC opposition. But below 40\% GC
content, large AT regions are possible that form bubbles freely,
hence dominating the bubble formation process.

\begin{figure}
\includegraphics[width=8.0cm]{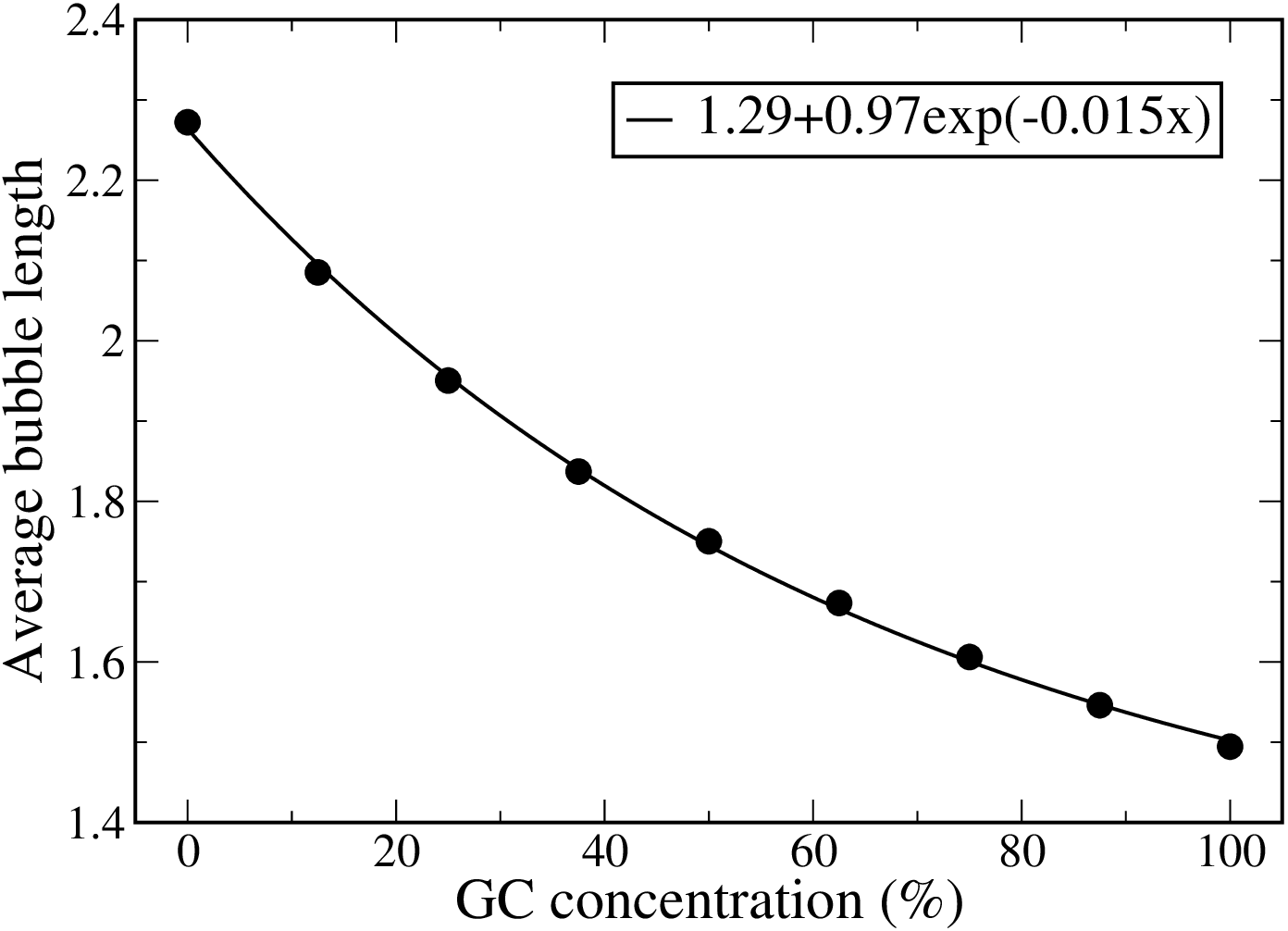}
\caption{\label{fig:3} Dependence of the average bubble length $L_B$,
Eq. (\ref{avl}), on the GC content of the sequence (points).
Continuous line shows a fitting with an exponential decay.
$T=310$ K, $y_{thres}=1.5$ \AA.}
\end{figure}

Figure \ref{fig:3} presents the average bubble length $L_B$, which is
given by the total number of base-pairs in bubble states divided by
the total number of bubbles:
\begin{equation} \label{avl}
L_B=\lim_{L\to\infty}\frac{ \sum_l l \langle N(l) \rangle}{ \sum_{l \geq 1}
\langle N(l) \rangle}=\frac{\sum_l lP(l)}{\sum_{l \geq 1} P(l)}.
\end{equation}
$L_B$ depends strongly on the GC content, showing an
exponential decay. This stresses the importance the sequence has on
the typical size of denaturation bubbles.

\begin{figure}
\includegraphics[width=8.0cm]{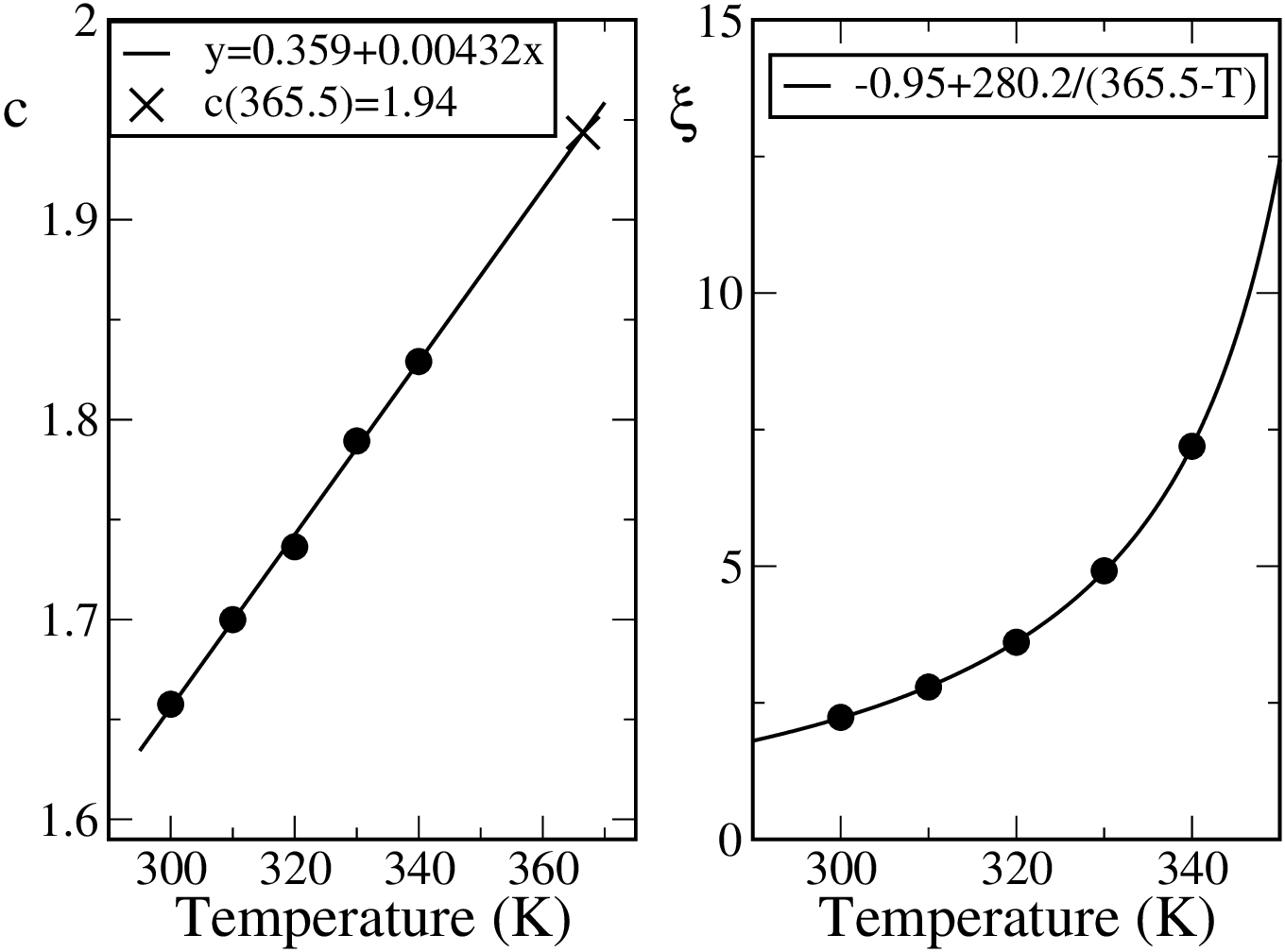}
\caption{\label{fig:4} Temperature dependence of the parameters of
distribution, Eq. (\ref{se}), for pure GC sequences.
Left: exponent $c$, the line is a linear fit.
Right: decay length $\xi$, the line is a fit with a divergent behavior as
$(T_c-T)^{-1}$, where the melting temperature is predicted
to be $T_c=365.5$ K, in agreement with transfer integral calculations.
}
\end{figure}

A recent work reports exponential decay of the bubble length
probability \cite{krastan}. Although this seems to
contradict our present results, this is not the case because the study
\cite{krastan} considers only a restricted regime of lengths,
extending from $l=3$ up to $l=12-14$.
Therefore, a restricted portion
of the distribution may appear as a seemingly straight line in a
semilogarithmic plot, indicating exponential decay. However, if one
looks for either smaller ($l=1, \; 2$) or larger lengths, below or
above this regime, the exponential law fails. On the contrary the
formula (\ref{se}), or a stretched exponential law \cite{preprint},
describes accurately the
whole regime from $l=1$ up to several tens of base-pairs. This means
that large bubbles are more probable than what it would have been
implied by exponential decay. Earlier studies of a simplified
version of the PBD model (using a linearized stacking interaction)
have also shown a slower than exponential
decay of the distribution function \cite{DPB2,sung}.

Importantly, the Poland-Scheraga (PS) model of DNA melting predicts a bubble
length distribution described by Eq.(\ref{se})
\cite{coluzzi,kafri,baiesi}. Hence our results establish a bridge
between these two different theoretical approaches, which could help to future
better understanding of the principles underlying the models, and henceforth a
more effective modeling approach to DNA.
In the PS model, the exponent $c$ has a very
important physical meaning, as its value at the critical temperature (where
$\xi$ diverges and the distribution is given by a pure power-law)
indicates the order of the melting transition:
for $c<1$ there is no phase transition, the transition is smooth for $1<c<3/2$,
second order for $3/2\leq c <2$ and first order for $c>2$.
It is interesting to characterize this exponent also for the PBD model;
work in this direction is in progress and a detailed temperature dependence
of bubble length distributions will be presented elsewhere \cite{us}.
Independently of the model, for $c>2$ the average bubble length
$L_B\propto\sum lP(l)$ remains finite
as $T\to T_c^{-}$, and it diverges for $c<2$, thus presenting a discontinuous or
a continuous transition, respectively.

As an advance of our work in progress, in Figure \ref{fig:4}
we depict a preliminary result for the
exponent $c$ and the decay length $\xi$
obtained for pure GC sequences at different
temperatures. As happens in the PS model \cite{coluzzi}, also here the
exponent increases linearly with temperature below the critical temperature.
The PS model also predicts that $\xi$ diverges as $(T_c-T)^{-1}$ \cite{coluzzi}.
Using this to fit the $\xi$ data, we obtain a value $T_c($GC$)=365.5$ K, in
agreement with transfer integral calculations we have performed.
Hence the
predicted value of the exponent $c$ at $T_c$ is $c=1.94$ (compatible with the
value $c=1.95$ used in biological studies \cite{yeramian}):
we would have a very sharp transition, but still not first order, if the
PS scheme applies. But we have to be very careful: for AT at $T=310$ K,
Figure \ref{fig:2} shows that $c=1.96$.
A transfer integral calculation yields $T_c($AT$)=325.5$ K, so the exponent at
$T_c$ is expected to be higher than 2. Work in this direction is necessary, but
the possibility to speculate that not only the transition temperature but also the
nature of the phase transition itself depends on the sequence, is too tempting
to let it go. The role that the parameters of the model play in this has
also to be carefully investigated.
Interestingly, previous work using the Peyrard-Bishop model without the
anharmonic stacking interaction ($\rho=0$)
yields a value of $c$ much lower than 2 \cite{sung}. Thus, regardless
whether the transition in the full PBD model is first order or not,
the analysis of bubble distributions shows once again \cite{DPB,DP}
that the anharmonic stacking interaction is responsible for the
sharp ($c\approx 2$) transition in the model.


We are not aware of any experimental technique able to track individual bubbles
in long sequences in order to study their size distributions. However, with
study of the probabilities of complete melting of molecules of different
sizes, the power law behavior we have described may appear. The
probability of complete melting of a given sequence has already been
experimentally measured \cite{zocchi} using a novel technique based on hairpin
quenching. This same technique, applied at constant temperature to sequences of
varying length but similar composition, should clarify how the formation of
bubbles depends on their size below the melting temperature.
However, it should be noted that this experiment measuring complete
melting is not exactly equivalent to our study of bubble formation:
sequence, finite-size and boundary effects
may play an important role.


In summary, using the Peyrard-Bishop-Dauxois model
we have shown that at physiological temperatures the formation
of thermally induced bubbles of different sizes follows a nonexponential
distribution with long tails, due to nonlinear interactions within base-pairs.
The occurrence of these openings may have an important effect,
which should be taken into account in biological processes involving the
opening of double stranded DNA. Moreover, structural properties of the DNA
molecule may also depend on the frequency of occurrence of these bubbles.


We acknowledge useful discussions with K.\O. Rasmussen,
N.K. Voulgarakis, M. Peyrard and A.R. Bishop.
S. A. acknowledges financial support from Ministerio de
Educaci\'on y Ciencia (Spain) through grant MOSAICO.



\begin{thebibliography}{99}
\bibitem{params} The parameter values are $K=0.025$ eV/\AA$^2$,
$\rho=2$, $b=0.35$ \AA$^{-1}$,
$D_{GC}=0.075$ eV, $a_{GC}=6.9$ \AA$^{-1}$ for a GC base pair, and
$D_{AT}=0.05$ eV, $a_{AT}=4.2$ \AA$^{-1}$ for an AT pair.
\bibitem{stexp} The fitting has been made using the least squares method with a
weight proportional to $(P(l))^{-2}$.
\bibitem{comment}
It has to be remarked that, although insightful and interesting,
the model with $\rho=0$ does not give a quantitatively correct description
of DNA melting \cite{DPB,DP}.
\end{thebibliography}
\end{document}